# The SDSS SkyServer – Public Access to the Sloan Digital Sky Server Data[1]

Alexander Szalay[1], Jim Gray[2],

Ani Thakar[1], Peter Z. Kunszt[4], Tanu Malik[1],

Jordan Raddick[1], Christopher Stoughton[3], Jan vandenBerg[1]

(1) The Johns Hopkins University,
(2) Microsoft,
(3) Fermi National Accelerator Laboratory, Batavia,
(4) CERN

{Szalay, Thakar, Raddick, Vincent}@pha.jhu.edu,
Gray@Microsoft.com,
Peter.Kunszt@cern.ch,
Stoughto@fnal.gov

---

[1] This article has been submitted for publication.

# The SDSS SkyServer – Public Access to the Sloan Digital Sky Server Data[2]

Alexander Szalay[1], Jim Gray[2], Ani Thakar[1], Peter Z. Kunszt[4], Tanu Malik[1],
Jordan Raddick[1], Christopher Stoughton[3], Jan vandenBerg[1]
(1) The Johns Hopkins University, (2) Microsoft, (3) Fermi National Accelerator Laboratory, Batavia, (4) CERN
Gray@Microsoft.com, {Szalay, Thakar, Raddick, Vincent}@pha.jhu.edu, Peter.Kunszt@cern.ch, Stoughto@fnal.gov

**Abstract:** *The SkyServer provides Internet access to the public Sloan Digital Sky Survey (SDSS) data for both astronomers and for science education. This paper describes the SkyServer goals and architecture. It also describes our experience operating the SkyServer on the Internet. The SDSS data is public and well-documented so it makes a good test platform for research on database algorithms and performance.*

## Introduction

The SkyServer provides Internet access to the public Sloan Digital Sky Survey (SDSS) data for both astronomers and for science education. The SDSS is a 5-year survey of the Northern sky (10,000 square degrees) to about ½ arcsecond resolution using a modern ground-based telescope [SDSS]. It will characterize about 200M objects in 5 optical bands, and will measure the spectra of a million objects. The first year's data is now public.

The raw data gathered by the SDSS telescope at Apache Point, New Mexico, is fed through data analysis software pipelines at Fermilab. Imaging pipelines analyze data from the imaging camera to extract about 400 attributes for each celestial object along with a 5-color "cutout" image. The spectroscopic pipelines analyze data from the spectrographs, to extract calibrated spectra, redshifts, absorption and emission lines, and many other attributes. These pipelines embody much of mankind's knowledge of astronomy [SDSS-EDR]. The pipeline software is a major part of the SDSS project: approximately 25% of the project's cost and effort. The result is a large high-quality catalog of the Northern sky, and of a small stripe of the Southern sky. When complete, the survey data will occupy about 40 terabytes (TB) of image data, and about 3 TB of processed data.

After calibration, the pipeline output is available to the astronomers in the SDSS consortium. After approximately a year, the SDSS publishes the data to the astronomy community and the public – so in 2007 all the SDSS data will be available to everyone everywhere.

The first year's SDSS data is now public. It is about 80GB containing about 14 million objects and 50 thousand spectra. You can access the public data via the SkyServer on the Internet (http://skyserver.sdss.org/) or you may get a private copy of the data. The web server supports both professional astronomers and educational access.

Amendments to the public SDSS data will be released as the data analysis pipeline improves, and the data will be augmented as more becomes public (next scheduled release is January 2003). In addition, the SkyServer will get better documentation and tools as we learn how it is used. There are Japanese and German versions of the website, and the server is being mirrored in many parts of the world.

This paper sketches the SkyServer design. It reports on the database and web site design, describes the data loading pipeline, and reports on early website usage.

## Web Server Interface Design

The SkyServer is accessed via the Internet using standard browsers. It accepts point-and-click requests for images of the sky, images of spectra, and for tabular outputs of the SDSS database. It also has links to the online literature about objects (e.g. NED, VizieR and Simbad). The site has an SDSS project description, tutorials on how the data was collected and what it means, and also has projects suitable to teach or learn astronomy and computational science at various grade levels. Figure 1 cartoons the main access screens.

The simplest and most popular access is a coffee-table atlas of *famous places* that shows color images of interesting (and often famous) celestial objects. These images try to lead the viewer to articles about these objects, and let them drill down to view the objects within the SDSS data. There are also tools that let the user navigate by field or plate to get images and spectra of particular objects (see Figure 1). To drill down further, there is a text and a GUI SQL interface that lets sophisticated users mine the SDSS database. A point-and-click pan-zoom scheme lets users pan across a section of the sky and pick objects and their spectra (if present).

The sky color images were built specially for the website. The original 5-color 80-bit deep images were converted using a nonlinear intensity mapping to reduce the brightness dynamic range to screen quality. The augmented-color images are 24bit RGB, stored as JPEGs. An image pyramid was built at 4 zoom levels. The spectra are also converted to 8bit GIF images.

[2] The Alfred P. Sloan Foundation, the Participating Institutions, the National Aeronautics and Space Administration, the National Science Foundation, the U.S. Department of Energy, the Japanese Monbukagakusho, and the Max Planck Society have provided funding for the creation and distribution of the SDSS Archive. The SDSS Web site is http://www.sdss.org/. The Participating Institutions are The University of Chicago, Fermilab, the Institute for Advanced Study, the Japan Participation Group, the Johns Hopkins University, the Max-Planck-Institute for Astronomy (MPIA), the Max-Planck-Institute for Astrophysics (MPA), New Mexico State University, Princeton University, the United States Naval Observatory, and the University of Washington. Compaq donated the hardware for the SkyServer and some other SDSS processing. Microsoft donated the basic software for the SkyServer.

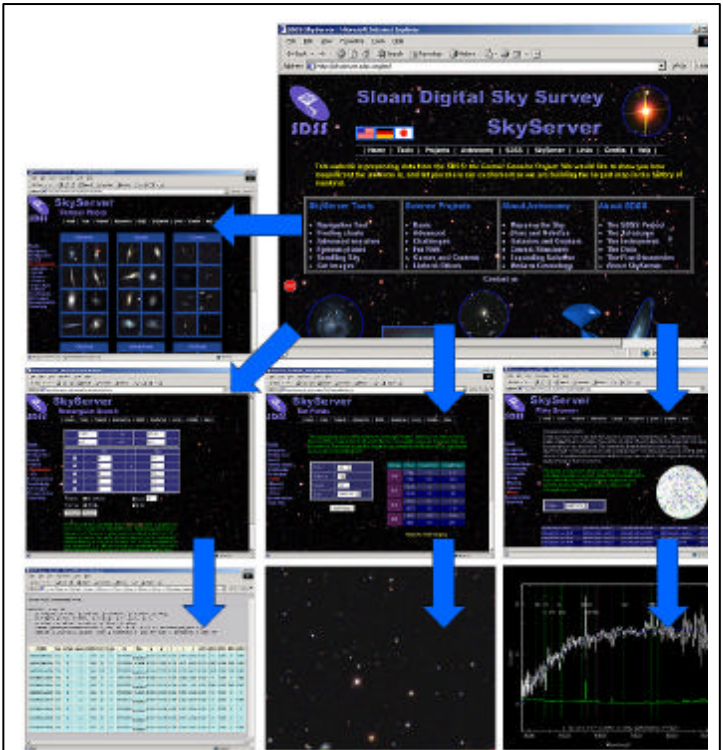

**Figure 1:** The SkyServer web interface provides many different ways to look at the SDSS data. The simplest is "famous places" which is just a gallery of beautiful images. More sophisticated users can navigate to find the images and the data for a particular celestial object. There are a variety of query interfaces and also links to the online literature about objects.

The SkyServer is just one of the ways to access the SDSS data. There is also the Catalog Archive Server (CAS) which is an ObjectivityDB™ database built by Johns Hopkins University (http://www.sdss.jhu.edu/ScienceArchive/). Much of the SkyServer database architecture is copied from the CAS database design to leverage its database documentation and design. In addition, the raw SDSS pixel-level files are available from Data Archive Server (DAS) at Fermilab (http://sdssdp7.fnal.gov/cgi-bin/das/main.cgi/). The CAS and DAS are operated by Fermilab and accessed via Space Telescope Science Institute's MAST (Multi Mission Archive at Space Telescope) website at http://archive.stsci.edu/sdss/

### SkyServer Data Mining

Data mining was our original motive in building the SQL-based SkyServer. We wanted a tool that would be able to quickly answer questions like: "find gravitational lens candidates" or "find other objects like this one." Indeed, we [Szalay] defined 20 typical queries and designed the SkyServer database to answer those queries. Another paper describes the queries and their performance [Gray], but we quickly summarize the results here.

The queries correspond to typical tasks astronomers would do with a C++ program, extracting data from the archive, and then analyzing it. Being able to state queries simply and quickly could be a real productivity gain for the Astronomy community. We were very pleased to discover that all 20 queries have fairly simple SQL equivalents. This was not obvious when we started. Often the query can be expressed as a single SQL statement. In some cases, the query is iterative, the results of one query feeds into the next.

Many of the queries run in a few seconds. Some involving a sequential scan of the database take about 3 minutes. A few complex joins take nearly an hour. Occasionally the SQL optimizer picks a very bad plan and a query can take several hours. The spatial data queries are both simple to state and execute quickly using a spatial index. We circumvented a limitation in SQL Server by pre-computing the neighbors of each object. Even without being forced to do it, we might have created this materialized view to speed queries. In general, the queries benefited from indices and column subsets containing popular fields.

Translating the queries into SQL requires a good understanding of astronomy, a good understanding of SQL, and a good understanding of the database. "Normal" astronomers use very simple SQL queries. They use SQL to extract a subset of the data and then analyze that data on their own system using their own tools. SQL, especially complex SQL involving joins and spatial queries, is just not part of the current astronomy toolkit. This stands as a barrier to wider use of the SkyServer by the astronomy community. We hope that a good visual query tool that makes it easier to compose SQL will ameliorate this problem.

### SkyServerQA-The SDSS Query Tool

SkyServerQA is a GUI SQL query tool to help compose SQL queries. It was inspired by the SQL Server Query Analyzer, but runs as a Java applet on UNIX, Macintosh, and Windows clients and is freely available from the SDSS web site [Malik]. It connects via ODBC/JDBC (for local use) and via HTTP or SOAP for use over the Internet.

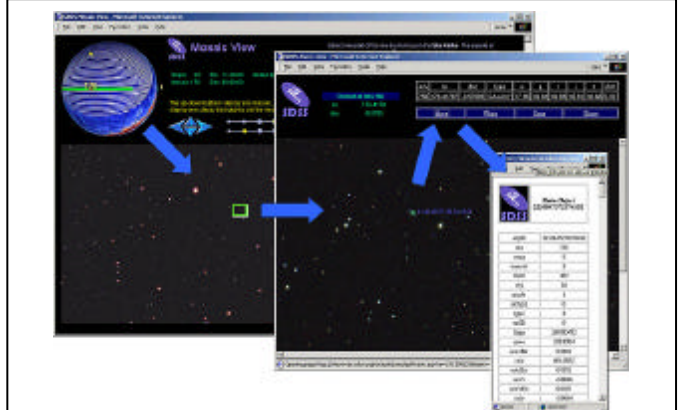

**Figure 2:** The navigation interface allows you to point to a spot on the celestial globe to view the "stripe" for that spot. Then you can zoom in 3 levels to view objects close up. By pointing to an object you can get a summary of its attributes from the database, and one can also call up the whole record.

SkyServerQA provides both a text-based and a diagram-based query mode. In the text-based mode, the user composes and executes SQL queries, stored procedures, or functions. The text based query window is shown on the left of Figure 3. In the diagram-based mode, the user formulates the query from icons, lists, and options in the left pane, without needing to know any syntax. While the user creates the query diagram, SkyServerQA creates the syntactically correct SQL query. This implicitly teaches SQL. The GUI is shown on the right of Figure 3.

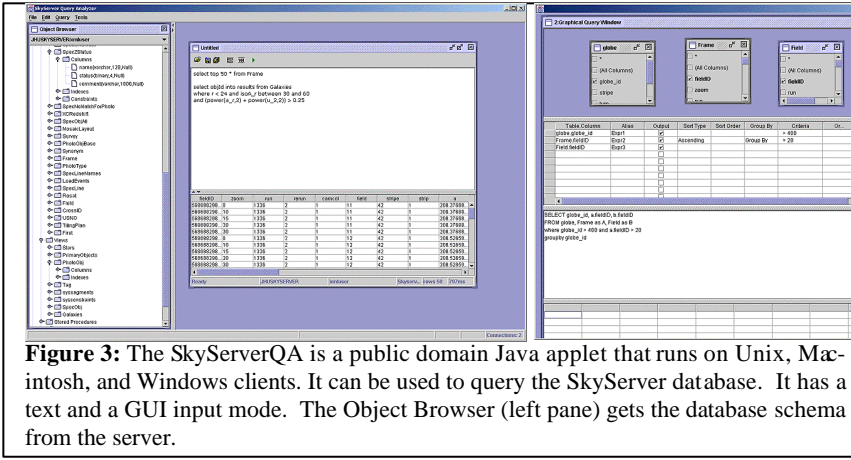

**Figure 3:** The SkyServerQA is a public domain Java applet that runs on Unix, Macintosh, and Windows clients. It can be used to query the SkyServer database. It has a text and a GUI input mode. The Object Browser (left pane) gets the database schema from the server.

SkyServerQA builds a hierarchical object browser of the database, tables, stored procedures, functions, columns, indexes, dependencies, and comments in the left pane from the database catalog (see Figure 3). Tool tip text tells the user the meaning of each table and field. Metadata includes data types, lengths, and null indicators. Indices consist of the columns on which they are built. Constraints show the Primary Key constraint for the table as well as Foreign Key constraints. Foreign Key constraints show the table to which they reference.

SkyServerQA provides results in three formats
1. Grid Based, for quick viewing of the results,
2. CSV, in ASCII comma separated values, for use in spreadsheets etc., and
3. XML, for use in any application that can read XML data

We plan to add FITS format as a fourth option [FITS]. The user can save these results to a file.

Query execution statistics are vital for large result-sets. The status window shows the execution time of each query, rounded to the nearest second. It also shows the connection information of the user, catalog name and server name.

The public SkyServer web server limits the queries to 1,000 records or 30 seconds of computation. For more demanding queries, the users must attach to a "private" SkyServer.

Once the query answer is produced, there is still a need to understand it. We have not made any progress on the data visualization problems posed in [Szalay].

**Web Server Design**

The SkyServer's architecture is fairly simple: a front-end IIS web server accepts HTTP requests processed by JavaScript Active Server Pages (ASP). These scripts use Active Data Objects (ADO) to query the backend SQL database server. SQL returns record sets that the JavaScript formats into pages.

The website is about 10,000 lines of JavaScript and was built by two people as a spare-time activity.

This design derives from the TerraServer [Barclay] – both the structured data and the images are all stored in the SQL database. A 4-level image pyramid of the frames and stripes is precomputed, allowing users to see an overview of the sky, and then zoom into specific areas for a close-up view of a particular object.

The most challenging aspect of the web site design has been supporting a rich user interface for many different browsers. Supporting Netscape Navigator™, Mozilla™, and Microsoft Internet Explorer™ is a challenge – especially when the many Windows™, Macintosh™, and UNIX™ variants are considered. The SkyServer does not download applets to the clients (except for SkyServerQA), but it does use both cascading style sheets and dynamic HTML. It is an ongoing struggle to support the browsers as they evolve.

Professional astronomers generally have a good command of English, but SkyServer supports an international user community that includes children and non-scientists. So, the web page hierarchy branches three ways: there is an English branch, a German branch, and a Japanese branch. Other languages can be added by people fluent in those languages. Each mirrored site will have all the data and supports all the languages.

**SkyServer used for Education**

The public access to real astronomical data and the SkyServer's web interfaces make it an enormous potential resource for science education and public outreach. Today, most students learn astronomy through textbook exercises that use artificial data or data that was taken centuries ago. With SkyServer, students can study data from galaxies never before seen by human eyes. We are designing several interactive educational projects that let students use SkyServer to learn concepts from astronomy and computational science.

We target the projects at two audiences: first, bright students excited about astronomy who want to work with data independently, and second, students taking general astronomy or other science courses as part of a school curriculum. To accommodate both audiences, we offer several different project levels, from "For Kids" (projects for elementary school students) to "Challenges" (projects designed to stretch bright college undergraduates). All projects designed for use in schools include a password-protected teachers' site with solutions, advice on how to lead classes through projects and correlations to national education standards [Project 2061].

For example, a kids' project, "Old Time Astronomy," (http://skyserver.sdss.org/en/proj/kids/oldtime/) asks students to imagine what astronomy was like before the cam-era was invented, when astronomers had to record data through sketches. The project shows SDSS images of stars and galaxies, and then asks students to sketch what they see. After a student has sketched the image, she trades with another student to see if the other student can guess which image was sketched (see Figure 4.)

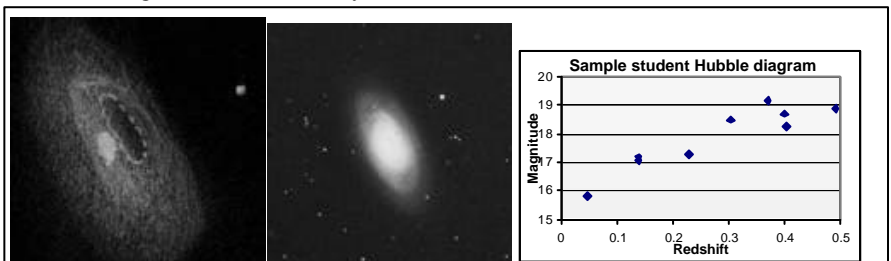

**Figure 4**: An example from the "Old Time Astronomy" project compares the sketch of Galaxy M64 made by amateur astronomer Michael Geldorp (left) to the image of the same galaxy from the Digitized Sky Survey (right). A more advanced project has students plot a Hubble diagram at right (showing redshift and relative distance of nine galaxies) to "discover" the expansion of the universe.

A project for advanced high school students and college undergraduates explores the expanding universe. The web site first gives students background reading about how scientists know the universe is expanding. Then, it lets students discover the expansion for themselves by making a *Hubble Diagram* – a plot of the velocities (or redshifts) of distant galaxies as a function of their distances from Earth. A sample student Hubble diagram is shown in Figure 4. Among other things, this teaches students how to work with real data.

Many more exercises and projects are being developed around the SkyServer. One particularly successful one was done by a teacher and some students in Mexico – there is growing international interest in using the SDSS to teach science to students in their native language (Spanish in that case).

One of the most exciting aspects of using SkyServer in education is its potential for students to pose and answer groundbreaking astronomical research questions. Because students can examine exactly the same data as professional astronomers, they can ask the same questions. Each school project ends with a "final challenge" that invites students to do independent follow-up work on a question that interests them. We are also working on a mentorship program that will match students working on school science fair projects with professional astronomers that volunteer to act as mentors, helping students to refine their research questions and to obtain the data they need to find answers.

### Site Traffic

The SkyServer has been operating since 5 June 2001. In the first 4 months it served about two million hits, 700 thousand page views via 50 thousand sessions. About 4% of these are to the Japanese sub-web and 3% to the German sub-web. The educational projects got about 8% of the traffic: about 250 page views a day. The server has been up 99.98% of the time. There have been 11 reboots, 8 to for software upgrades and 3 associated with failing power. The patches cause outages of 5 minutes, the power and operations outages last several hours. Not shown in the statistics, but clearly visible in Figure 5 are two network outages or overloads that plagued Fermilab on 22 June and 26 July. Conversely, the peak traffic coincided with classes using the site, news articles mentioning it, or with demonstrations at Astronomy conferences. The sustained usage is about 400 people accessing about 3,000 pages per day. The site is configured to handle a load 100x larger than that. A TV show on October 2, generated a peak 20x the average load. About 30% of the traffic is from other sites "crawling" the SkyServer -- extracting the data and images. There are about 10 "hacker attacks" per day.

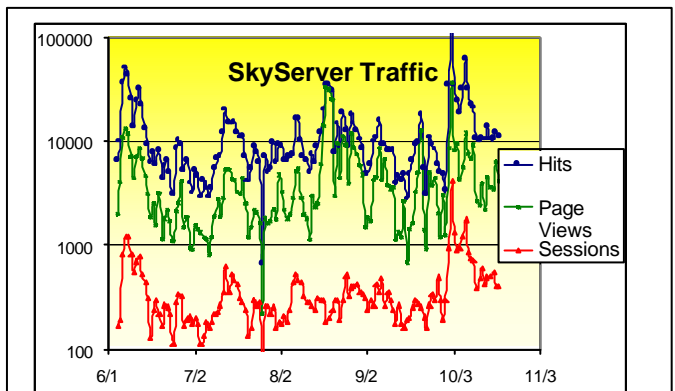

**Figure 5:** In 4 months the SkyServer processed about 2 million page hits, about ½ a million pages and about thirty thousand sessions.

### Web Server Deployment and Administration

The application is primarily administered from Johns Hopkins and San Francisco using the Windows™ remote windows system (Terminal Server) feature. The Fermilab staff manages the physical hardware, the network, and site security. There is a mirror server at Johns Hopkins for incremental development and testing. The two sites are synchronized about once per week.

**The Sloan Digital Sky Survey Data and Databases**

The SDSS processing pipeline at Fermilab examines the 5-color images from the telescope and identifies ***photo objects*** as either *stars*, *galaxies*, *trail* (cosmic ray, satellite,…), or some *defect*. The classification is probabilistic — i.e., it is sometimes difficult to distinguish a faint star from a faint distant small galaxy. In addition to the basic classification, the pipeline extracts about 400 attributes from an object, including a "cutout" of the object's pixels in the 5 color bands.

The actual observations are taken in stripes about 2.5° wide and 120° long (see Figure 6). To further complicate things, these stripes are in fact the mosaic of two night's observations (two strips) with about 10% overlap. The stripes themselves have some overlaps near the horizon. Consequently, about 11% of the objects appear more than once in the pipeline. The pipeline picks one object instance as *primary* but all instances are recorded in the database. Even more challenging, one star or galaxy often overlaps another, or a star is part of a cluster. In these cases *child* objects are *deblended* from the parent object, and each child also appears in the database (deblended parents are never primary.) In the end about 80% of the photo objects are primary.

The photo objects have positional attributes - right ascension and declination in the J2000 coordinate system, also represented as the Cartesian components of a unit vector, and an index into a Hierarchical Triangular Mesh (HTM). They also have brightness stored in logarithmic units (magnitudes) with error bars in each of the five color bands. These magnitudes are measured in six different ways (for a total of 60 attributes). The image processing pipeline also measures each galaxy's extent in several ways in each of the 5 color bands with error estimates. The pipeline assigns about a hundred additional properties to each object – these attributes are variously called flags, status, and type and are encoded as bit flags.

The pipeline correlates each object with objects in other catalogs: United States Naval Observatory [USNO], Röntgen Satellite [ROSAT], Faint Images of the Radio Sky at Twenty-centimeters [FIRST], and others. Successful correlations are recorded in a set of relationship tables.

Spectrograms are the second kind of main data product produced by the Sloan Digital Sky Survey. About 600 spectra are observed at once using a single plate with optical fibers going to different CCDs. The pipeline processing typically extracts about 30 spectral lines from each spectrogram and carefully estimates the object's redshift.

**The Relational Database Design**

Originally, the SDSS developed the entire database on ObjectivityDB™ [SDSS-Science Archive]. The designers used subclasses extensively: for example the PhotoObject has Star and Galaxy subclasses. ObjectivityDB supports arrays so the 5-colors naturally mapped to vectors of 5 values. Connections to parents, children, spectra, and to other surveys were represented as object references. Translating the ObjectivityDB™ design into a relational schema was not straightforward; but we wanted to preserve as much of the original design as possible in order to preserve the existing knowledge, skills, and documentation.

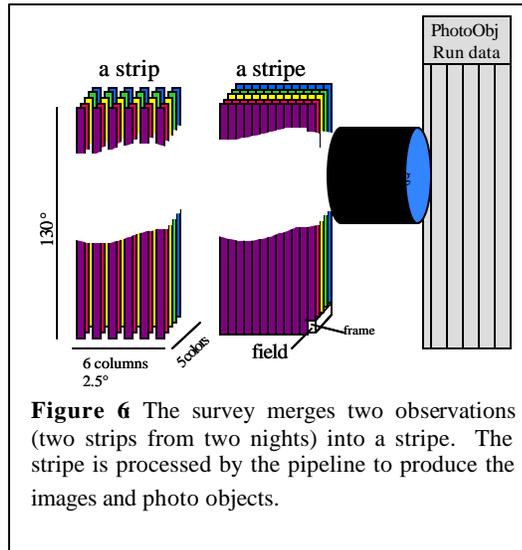

**Figure 6** The survey merges two observations (two strips from two nights) into a stripe. The stripe is processed by the pipeline to produce the images and photo objects.

The SQL relational database language does not support pointers, arrays, or sub-classing – it is a much simpler data model. This is both a strength and a liability. We approached the SQL database design by using views for subclassing, and by using foreign keys for relationships.

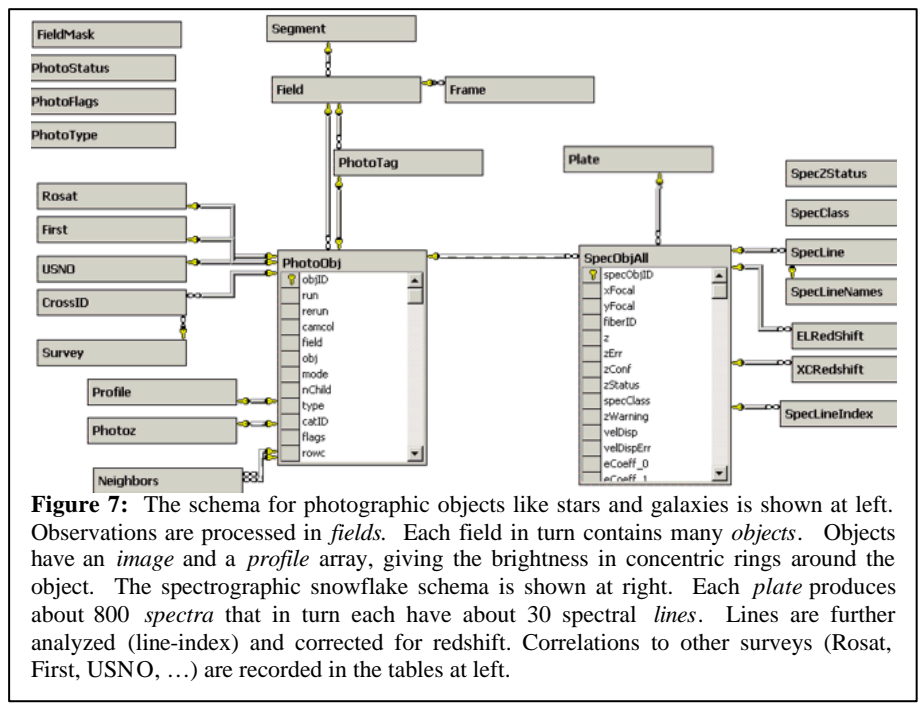

**Figure 7:** The schema for photographic objects like stars and galaxies is shown at left. Observations are processed in *fields*. Each field in turn contains many *objects*. Objects have an *image* and a *profile* array, giving the brightness in concentric rings around the object. The spectrographic snowflake schema is shown at right. Each *plate* produces about 800 *spectra* that in turn each have about 30 spectral *lines*. Lines are further analyzed (line-index) and corrected for redshift. Correlations to other surveys (Rosat, First, USNO, …) are recorded in the tables at left.

## Photographic Objects

Starting with the imaging data, the `PhotoObj` table has all the star and galaxy attributes. The 5 color attribute arrays and error arrays are represented by their names (u, g, r, i, z.) For example, `ModelMag_r` is the name of the "red" magnitude as measured by the best model fit to the data. In cases where names were unnatural (for example in the profile array) the data is encapsulated by access functions that extract the array elements from a blob. Pointers and relationships are represented by "foreign keys".

Views are used for sub-classing. The *primaryObjects*, *galaxies*, and *stars*, subclasses of the `PhotoObj` class are defined as the `PrimaryObjects, Stars, and Galaxies` views of the `PhotoObj` base table (views are virtual tables that just translate into queries on the base table).

The result is a snow-flake schema with the `photoObj` table in the center other tables clustered about it (Figure 7). The 14 million `photoObj` records each have about 400 attributes describing the object – about 2KB per record. The `Field` table describes the processing that was used for all objects in that field, in all frames. The other tables connect the *photoObj* to literals (e.g. `flags & fPhotoFlags('primary')`), or connect the object to objects in other surveys. One table, `neighbors`, is computed after the data is loaded. For every object the *neighbors* table contains a list of all other objects within ½ arcminute of the object (typically 10 objects). This speeds proximity searches.

## Spectroscopic Objects

Spectrograms are the second kind of data object produced by the Sloan Digital Sky Survey. About 600 spectra are observed at once using a single plate with optical fibers going to two different spectrographs. The plate description is stored in the `plate` table, and the description of the spectrogram is stored in the *specObj* table (each spectrogram has a handsome GIF image associated with it.). The pipeline processing typically extracts about 30 spectral lines from each spectrogram. The spectral lines are stored in the `SpecLine` table. The `SpecLineIndex` table has quantities derived from analyzing line groups. These quantities are used by astronomers to characterize the types and ages of astronomical objects. Each line is crosscorrelated with a model and corrected for redshift. The resulting attributes are stored in the `xcRedShift` table. A separate redshift is derived using only emission lines. Those quantities are stored in the `elRedShift` table. Again, as is standard with relational database designs, all these tables are integrated with foreign keys – each specObj object has an unique `specObjID` key, and that same key value is stored as part of the key of every related spectral line. To find all the spectral lines of object 512 one writes the query

```
select *
from specLine
where specObjID = 512
```

The spectrographic tables also form a snowflake schema that gives names for the various flags and line types. Foreign keys connect `PhotoObj` and `SpecObj` tables if a photo object has a measured spectrogram.

There are also a set of "miscellaneous" tables used to monitor the data loading process and to support the web interface.

## Database Access Design – Views, Indices, and Access Functions

The *photoObj* table contains many types of objects (primaries, secondaries, stars, galaxies,…). In some cases, users want to see all the objects, but typically, users are just interested in primary objects (best instance of a deblended child), or they want to focus on just Stars, or just Galaxies. So, three views are defined on the *PhotoObj* table:
   *PrimaryObjects*: *photoObj* with flags('primary')=true
   *Stars*: *PrimaryObjects* with type='star'
   *Galaxies*: *PrimaryObjects* with type='galaxy'

Most users will work in terms of these tables rather than the base table. In fact, most queries are cast in terms of these views. This is the equivalent of sub-classing. The SQL query optimizer rewrites such queries so that they map down to the base *photoObj* table with the additional qualifiers.

To speed access, the *photoObj* table is heavily indexed (these indices also benefit the views). In a previous design based on an object-oriented database ObjectivityDB™ [Thakar], the architects replicated vertical data slices, called *tag* tables, which contain the most frequently accessed object attributes. These tag tables are about ten times smaller than the base tables (a few hundred1 bytes rather than a few thousand bytes).

Our concern with the tag table design is that users must know which attributes are in a tag table and must know if their query is *covered* by the fields in the tag table. Indices are an attractive alternative to tag tables. An index on fields A, B, and C gives an automatically managed tag table on those 3 attributes plus the primary key – and the SQL query optimizer automatically uses that index if the query is covered by (contains) only those 3 fields. So, indices perform the role of tag tables and lower the intellectual load on the user. In addition to giving a column subset, thereby speeding access by ten to one hundred fold, indices also cluster data so that searches are limited to just one part of the object space. The clustering can be by type (star, galaxy), or space, or magnitude, or any other attribute. One limitation is that Microsoft's SQL Server 2000 limits indices to 16 columns.

Today, the SkyServer database has tens of indices, and more will be added as needed. The nice thing about indices is that when they are added, they speed up any queries that can use them. The downside is that they slow down the data insert process – but so far that has not been a problem. About 30% of the SkyServer storage space is devoted to indices.

Nonetheless, we created a *PhotoTag* table for compatibility with the ObjectivityDB™ design, and to address the 16-column index limit of SQL Server 2000.

In addition to the indices, the database design includes a fairly complete set of foreign key declarations to insure that every profile has an object; every object is within a valid field, and so on. We also insist that all fields are non-null. These integrity constraints are invaluable tools in detecting errors during loading and they aid tools that automatically navigate the database.

**Spatial Data Access**

The SDSS scientists are especially interested in the galactic clustering and large-scale structure of the universe. In addition, the web interface routinely asks for all objects in a certain rectangular or circular area of the celestial sphere.

The SkyServer uses three different coordinate systems. First right-ascension and declination (comparable to latitude-longitude in terrestrial coordinates) are ubiquitous in astronomy. The (x,y,z) unit vector in J2000 coordinates is stored to make arc-angle computations fast. The dot product and the Cartesian difference of two vectors are quick ways to determine the arc-angle or distance between them.

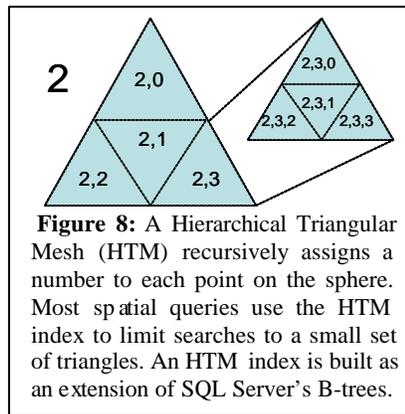

**Figure 8:** A Hierarchical Triangular Mesh (HTM) recursively assigns a number to each point on the sphere. Most spatial queries use the HTM index to limit searches to a small set of triangles. An HTM index is built as an extension of SQL Server's B-trees.

To make spatial area queries run quickly, the Johns Hopkins *hierarchical triangular mesh* (HTM) code [HTM, Kunszt1] was added to SQL Server. Briefly, HTM inscribes the celestial sphere within an octahedron and projects each celestial point onto the surface of the octahedron. This projection is approximately iso-area.

HTM partitions the sphere into the 8 faces of an octahedron. It then hierarchically decomposes each face with a recursive sequence of triangles – so each level of the recursion divides each triangle into 4 sub-triangles (Figure 8). SDSS uses a 20-deep HTM so that the individual triangles are less than 0.1 arcseconds on a side. So, the HTM ID for a point very near the north pole (in galactic coordinates) would be something like 3,0,….,0. There are basic routines to convert between (ra,dec) and HTM coordinates.

These HTM IDs are encoded as 64-bit integers. Importantly, all the HTM IDs within the triangle 6,1,2,2 have HTM IDs that are between 6,1,2,2 and 6,1,2,3. So, when the HTM IDs are mapped into a B-tree index they provide quick index for all the objects within a given triangle. The HTM library is an SQL extended stored procedure wrapped in a table-valued function spHTM_Cover(<area>). The <area> can be either a circle (ra, dec, radius), a half-space (the intersection of planes), or a polygon defined by a sequence of points. The function returns a table, each row defining the start and end of an HTM triangle. One can join this table with the PhotoObj table to get a spatial subset of photo objects.

The spHTM_Cover() function is too primitive for most users, they actually want the objects nearby a certain object, or they want all the objects in a certain area. So, simpler functions are also supported. For example: fGetNearestObjEq(1,1,1) returns the nearest object within one arcminute of equatorial coordinate (1º, 1º). These procedures are frequently used in queries and in the website access pages.

**Summary of Database Design**

In summary, the logical database design consists of photographic and spectrographic objects. They are organized into a pair of snowflake schemas. Subsetting views and many indices give convenient access to the conventional subsets (stars, galaxies, ...). Procedures and indices are defined to make spatial lookups convenient and fast.

**Database Physical Design**

The SkyServer initially took a simple approach to database design – and since that worked, we stopped there. The design counts on the SQL storage engine and query optimizer to make all the intelligent decisions about data layout and data access.

The data tables are all created in one file group. The database files are spread across 4 mirrored volumes. Each of the 4 volumes holds a database file that starts at 20 GB and automatically grows as needed. The log files and temporary database are also spread across these disks. SQL Server stripes the tables across all these files and hence across all these disks. It detects the sequential access, creates the parallel prefetch threads, and uses multiple processors to analyze the data as quickly as the disks can produce it. When reading or writing, this automatically gives the sum of the disk bandwidths (up to 140 MBps) without any special user programming.

**Table 1**: Count of records and bytes in major tables. Indices approximately double the space.

| Table | Records | Bytes |
|:---:|:---:|:---:|
| Field | 14k | 60MB |
| Frame | 73k | 6GB |
| PhotoObj | 14m | 31GB |
| Profile | 14m | 9GB |
| Neighbors | 111m | 5GB |
| Plate | 98 | 80KB |
| SpecObj | 63k | 1GB |
| SpecLine | 1.7m | 225MB |
| SpecLineIndex | 1.8m | 142MB |
| xcRedShift | 1.9m | 157MB |
| elRedShift | 51k | 3MB |

Beyond this file group striping, SkyServer uses all the SQL Server default values. There is no special tuning. This is the hallmark of SQL Server – the system aims to have "no knobs" so that the out-of-the box performance is quite good. The SkyServer is a testimonial to that goal. A later section discusses the hardware and the system performance.

**Database Load Process**

The SkyServer is a data warehouse: new data is added in batches, but mostly the data is queried. Of course these queries create intermediate results and may deposit their answers in temporary tables, but the vast bulk of the data is read-only.

Occasionally, a new schema is loaded (we are on V3 right now), so the disks were chosen to be large enough to hold three complete copies of the database.

From the SkyServer administrator's perspective, the main task is data loading -- which includes data validation. When new photo objects or spectrograms come out of the pipeline, they have to be added to the database. We are the system administrators – so we wanted this loading process to be as automatic as possible.

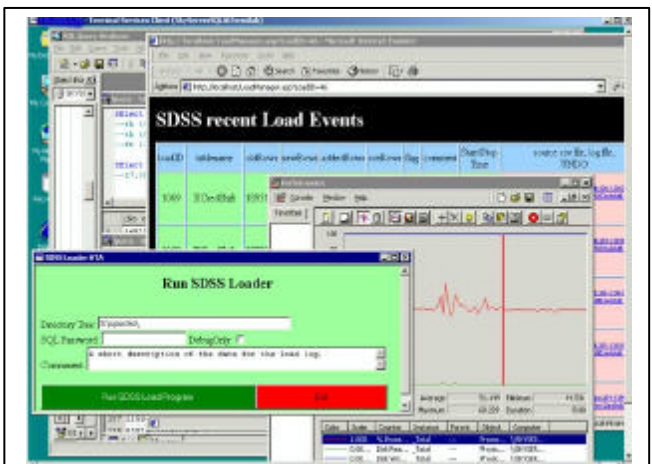

**Figure 9:** A screen shot of the SkyServer Database operations interface. The SkyServer is operated via the Internet using Winodows2000 Terminal Server, a remote desktop facility built into the operating system. Both loading and software maintenance are done in this way. This screen shot shows a window into the backend system after a load step has completed. It shows the loader utility, the load monitor, a performance monitor window and a database query window. This remote operation has proved a godsend, allowing the Johns Hopkins, Microsoft, and Fermilab participants to perform operations tasks from their offices.

The SDSS data pipeline produces FITS files, but also produces comma-separated list (csv) files of the object data and PNG files. The PNG files are converted to JPEG at various zoom levels, and an image pyramid is built before loading. These files are then copied to the SkyServer. From there, a script loads the data using the SQL Server's Data Transformation Service. DTS does both data conversion and the integrity checks. It also recognizes file names in some fields, and uses those names to place the contents of the corresponding image file (JPEG) as a blob field of the record. There is a DTS script for each table load step. In addition to loading the data, these DTS scripts write records in a *loadEvents* table recording the load time, the number of records in the source file, and the number of inserted records. The DTS steps also write trace files indicating the success or errors in the load step. A particular load step may fail because the data violates foreign key constraints, or because the data is invalid (violates integrity constraints.) In the web interface helps the operator to (1) undo the load step, (2) diagnose and fix the data problem, and (3) re-execute the load on the corrected data.

Loading runs at about 5 GB per hour (data conversion is very cpu intensive), so the current SkyServer data loads in about 12 hours.

A simple web user interface displays the load-events table and makes it easy to examine the CSV file and the load trace file. If the input file is easily repaired, that is done by the administrator, but often the data needs to be regenerated. In either case the first step is to UNDO the failed load step. Hence, the web interface has an UNDO button for each step.

The UNDO function works as follows: Each table in the database has a timestamp field that tells when the record was inserted (the field has Current_Timestamp as its default value.) The load event record tells the table name and the start and stop time of the load step. Undo consists of deleting all records of that table with an insert time between the bad load step start and stop times.

**Personal SkyServer**

A 1% subset of the SkyServer database (about .5 GB SQL Server database) can fit on a CD or be downloaded over the web. This includes the web site and all the photo and spectrographic objects in a 6º square of the sky. This personal SkyServer fits on laptops and desktops. It is useful for experimenting with queries, for developing the web site, and for giving demos. Essentially, any classroom can have a mini-SkyServer per student. With disk technology improvements, a large slice of the public data will fit on a single disk by 2003.

**Hardware Design and Performance**

The SDSS early data release database is about 60 GB (see table 1). It can run on a single processor system with just one disk, but the SkyServer at Fermilab runs on more capable hardware generously donated by Compaq Computer Corporation. Figure 10 shows the hardware configuration.

The web server runs Windows2000 on a Compaq™ DL380 with dual 1GHz PentiumIII processors. It has 1GB of 133MHz SDRAM, a 64-Bit/66MHz Single Channel Ultra3 SCSI Adapter with a mirrored disk. This web server does almost no disk IO during normal operation, but we clocked the disk subsystem at over 30MBps. The web server is also a firewall, it does not do routing and it has a separate "private" 100Mbps Ethernet link to the backend database server.

Most data mining queries are IO-bound, so the database server is configured to give fast sequential disk bandwidth, healthy CPU power, and high availability. The database server is a Compaq ProLiant ML530 running SQL Server 2000 and Windows2000. It has two 1GHz Pentium III Xeon processors, 2GB of 133MHz SDRAM; a 2-slot 64bit/66MHz PCI bus, a 5-slot 64bit/33MHz PCI, and a 32bit PCI bus. It has two 64-Bit/66MHz Single Channel Ultra3 SCSI Adapters. The ML530 also has a complement of high-availability hardware components: redundant, hot-swappable power supplies, redundant, hot-swappable fans, and hot-swappable SCA-2 SCSI disks.

The server has ten Compaq 37GB 10K rpm Ultra160 SCSI disks, five on each SCSI channel. Windows2000's software RAID manages these as five mirrors (RAID1's): one for the operating system and software, and the remaining four for databases. The database filegroups (both data and log) are spread across these four mirrors. SQL Server stripes the data across the four volumes, effectively managing the data disks as a RAID10 (striping plus mirroring). This configuration can scan data at 140 MBps for a simple query like:
```
select count(*)
from photoObj
where (r-g)>1.
```

So, how well does this work? A separate paper gives detailed timings on twenty complex queries [Gray], but to summarize: a typical index lookup runs primarily in memory and completes within a second or two. SQL Server uses available memory to cache frequently used data. Index scans of the 14M row photo table run in 7 seconds "warm" (5 m records per second when cpu bound), and 17 seconds cold, on a 4-disk 2-cpu Server. Queries that scan the entire 30GB *photoObj* table run at about 140 MBps and so take about 3 minutes. These scans use the available CPUs and disks in parallel – but are IO bound.

When the SkyServer project began, the existing software was delivering 0.5 MBps and heavy CPU consumption – indeed that was the main reason for considering a relational database over ObjectivityDB™. The SkyServer goal was 50MBps at the user level on a single machine. Beyond that, arrays of parallel database machines could deliver more bandwidth. This would give 5-second response to simple queries, and 5-minute response to full database scans. As it stands SQL Server and the Compaq hardware exceeded the performance goals by 500%. In general 4-disk workstation-class machines run at the 70MBps PCI(32/33) bus limit, while 8-disk RAID0 server-class machines run at 350 MBps. As the SDSS data grows, arrays of more powerful machines should allow the SkyServer to return most answers within seconds or minutes depending on whether it is an index search, or a full-database scan.

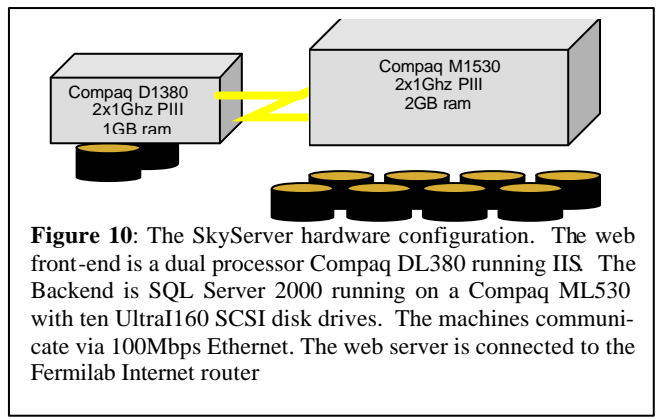

**Figure 10**: The SkyServer hardware configuration. The web front-end is a dual processor Compaq DL380 running IIS. The Backend is SQL Server 2000 running on a Compaq ML530 with ten UltraI160 SCSI disk drives. The machines communicate via 100Mbps Ethernet. The web server is connected to the Fermilab Internet router

We tried various disk-controller configurations to test maximum IO speed (see Figure 11):
- A 12-disk 4-controller software RAID0 system can scan at 430 megabytes per second with 12% cpu utilization at the file level [memspeed].
- SQL Server runs a minimal (`select count(*)`) database query at 331MBps on this hardware and also on nine disks on three controllers.
- At that rate, SQL is evaluating 2.6 million 128-byte tag records per second (mrps).
- At that rate for `count(*)` the cpu utilization is 75% -- 10 clocks per byte (cpb), 1300 clocks per record (cpr).
- The more complex query `count(*) where (r-g)>1` is cpu bound, using 19 cpb and 2300 cpr.
- When the data is all in main memory, the SQL System scans the data (`select count(*)`) at 5 mrps.
- A disk delivers up to 40 MBps. Three disks on one ultra3 controller deliver up to 119 MBps (99.8% scale up.)
- Three disks nearly saturate an ultra3 controller; beyond that additional controllers are needed.
- A 64-bit/33MHz PCI bus saturates at about 220MBps – not quite enough for two ultra3 controllers.

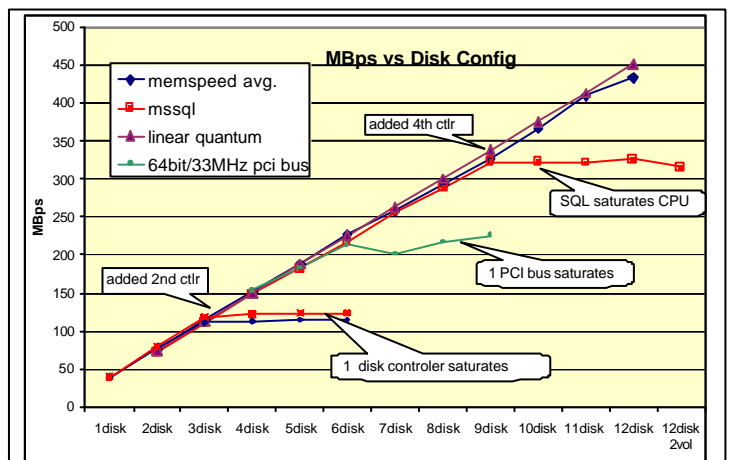

**Figure 11**: Sequential IO speed is important for data mining queries. This graph shows the sequential scan speed (megabytes per second) as more disks and controllers are added (one controller added for each 3 disks). It indicates that the SQL IO system process about 320MBps (and 2.7 million records per second) before it saturates.

## Summary and Next Steps

The SkyServer is a web site and SQL database containing the Sloan Digital Sky Survey early data release (about 14 million celestial objects and 50 thousand spectra.) It is accessed over the Internet via a website that provides point-and-click access as well as several query interfaces including form-based reports, raw SQL queries, and a GUI query builder.

We designed the SkyServer both as a website for easy public access and as a data mining site. To test the data mining capabilities; we implemented 20 complex astronomy queries and evaluated their performance. We are pleased with the performance. Others can clone the server for a few thousand dollars. The data and web server are also being replicate at various astronomy and computer science institutes to either explore the data or to experiment with better ways to analyze and visualize the data.

The tools described here are a modest step towards the goal of providing excellent data analysis and visualization tools. We hope that the SkyServer will enable computer scientists to advance this field by building better tools. The SkyServer now has four main evolutionary branches:

**Public SkyServer**: New versions of the SDSS data will be released once or twice a year as the data analysis pipeline improves, and the data will be augmented as more becomes public. In addition, the SkyServer will get better documentation and tools as we get more experience with how it is used. There are Japanese and German clones of the website, and there are plans to clone it at several other sites. A basic server costs less than ten thousand dollars.

**Virtual Observatory (VO)**: The SDSS is just one of many astronomy archives and resources on the Internet. The Internet will soon be the world's best telescope. It will have much of the world's astronomy data online covering all the measured spectral bands. The data will be cross-correlated with the literature. It will be accessible to everyone everywhere. And, if the VO is successful, there will be great tools to analyze all this data. The SkyServer is being federated with VirtualSky and NED at CalTech [VirtualSky] as Web Services, and with VizieR and Simbad at Strasbourg [VizieR, Simbad]. These are just first steps to a broader federation.

**The Science Archive** : A second group of SkyServers with preliminary (not yet released) data will form a virtual private network accessible to the SDSS consortium. These servers will have more sophisticated users who are intimately familiar with the data. So these servers will have unique needs.

**Outreach and Curriculum Development**: The SDSS data is a great vehicle for teaching both astronomy and computational science. The data is real -- everything comes with error bars, everything has a strong science component. The SDSS data also has strong graphical, spatial, and temporal components. It is fairly well documented and is public. And of course, it's big by today's standards. We hope that educators will "discover" the SkyServer and its educational potential – both at K-12 and at the university level.

## Acknowledgements

We acknowledge our obvious debt to the people who built the SDSS telescope, those who operate it, those who built the SDSS processing pipelines, and those who operate the pipeline at Fermilab. The SkyServer data depends on the efforts of all those people. Compaq and Microsoft generously donated hardware, software, and money to support the SkyServer. Tom Barclay advised us on the web site design, construction, and operation. Roy Gal, Steve Kent, Rich Kron, Robert Lupton, Steve Landy, Robert Sparks, Mark Subba Rao, Don Slutz, and Tamas Szalay contributed to the site's content and development. Sadanori Okamura, Naoki Yasuda, and Matthias Bartelmann built the Japanese and German versions of the site. Rosa Gonzalez and Kausar Yasmin helped with testing and developed some class martial.

## References

[Barclay] T. Barclay, D.R. Slutz, J. Gray, "TerraServer: A Spatial Data Warehouse," Proc. ACM SIGMOD 2000, pp: 307-318, June 2000
[FIRST] Faint Images of the Radio Sky at Twenty-centimeters (FIRST) http://sundog.stsci.edu
[FITS] FITS - Flexible Image Transport System, http://heasarc.gsfc.nasa.gov/docs/heasarc/fits.html
[Gray] SDSS 20 queries Answered. http://skyServer.sdss.org/en/download/
[HTM] Hierarchical Triangular Mesh http://www.sdss.jhu.edu/htm/
[Kunszt] P. Z. Kunszt, A. S. Szalay, I. Csabai, A. R. Thakar "The Indexing of the SDSS Science Archive" in Proc ADASS IX, eds. N. Manset, C. Veillet, D. Crabtree, (ASP Conference series), **216**, 141-145 (2000)
[Malik] SkyServer Query Analyzer, http://skyServer.sdss.org/en/download/
[MAST] Multi Mission Archive at Space Telescope. http://archive.stsci.edu/index.html
[Memspeed] http://research.microsoft.com/BARC/ Sequetial_IO /memspeed.zip
[NED] *NASA/IPAC Extragalactic Database*, http://nedwww.ipac.caltech.edu/
[Project 2061] Principles and Standards, http://www.project2061.org/
[ROSAT] Röntgen Satellite (ROSAT) http://heasrc.gsfc.nasa.gov/docs/rosat/rass.html
[SDSS] D.G. York et al., The Sloan Digital Sky Survey: Technical Summary, The Astronomical Journal. 120 (2000) 1579-1587,
[SDSS-EDR] C. Stoughton et. al., "Sloan Digital Sky Survey Early Data Release", The Astronomical Journal, in press (2002)
[SDSS-Science Archive] http://www.sdss.jhu.edu/ScienceArchive/doc.html
[Simbad] *SIMBAD Astronomical Database,* http://simbad.u-strasbg.fr/
[Szalay] A. Szalay, P. Z. Kunszt, A.Thakar, J. Gray, D. R. Slutz. "Designing and Mining Multi-Terabyte Astronomy Archives: The Sloan Digital Sky Survey," Proc. ACM SIGMOD 2000, pp.451-462, 2000
[Thakar] A.R. Thakar, P.Z. Kunszt, A.S. Szalayand G.P. Szokoly: "Multi-threaded Query Agent and Engine for a Very Large Astronomical Database,," in Proc ADASS IX, eds. N. Manset, C. Veillet, D. Crabtree, (ASP Conference series), **216**, 231 (2000).
[USNO] US Naval Observatory http://www.usno.navy.mil/products.shtml,
[Virtual Sky] *Virtual Sky*, http://VirtualSky.org/
[VizieR] *VizieR Service*, http://vizier.u-strasbg.fr/viz-bin/VizieR